\documentclass[final,1p,times]{elsarticle} 

\usepackage{graphicx}
\usepackage{amssymb} 
\usepackage{amsthm} 
\usepackage{lineno}
\usepackage{epstopdf}

\AppendGraphicsExtensions{.gif}

\newcommand{\pt}{$p_{T}$}

\newcommand{\auau}{Au+Au}
\newcommand{\gevc}{GeV/$c$}
\newcommand{\saaTwoHundred}{$\sqrt{s_{NN}} = 200$ GeV}
\newcommand{\saaSixtyTwo}{$\sqrt{s_{NN}} = 62.4$ GeV}

\journal{Nuclear Physics A} 

\begin{document}

\begin{frontmatter} 


\title{Measurements of Non-photonic Electron Production and Azimuthal
Anisotropy in $\sqrt {s_{NN}} = 39$, $62.4$ and $200$ GeV \auau\ Collisions from
STAR at RHIC}

\author{Mustafa Mustafa (for the STAR\fnref{col1} Collaboration)}
\fntext[col1] {A list of members of the STAR Collaboration and acknowledgements
can be found at the end of this issue.}
\address{Dept. Of Physics, Purdue University, West
Lafayette, IN 47907. US. E-mail mustafa@purdue.edu}

\begin{abstract} 
During the RHIC 2010 run, STAR has collected a large amount of minimum-bias, central and
high \pt\ trigger data in Au+Au collisions at $\sqrt{s_{NN}} = 39$, $62.4$
and 200 GeV with the detector configured to minimize photon conversion background. In this article we report on a new high precision measurement of
non-photonic electron mid-rapidity invariant yield, improved nuclear
modification factor and $v_{2}$ in Au+Au collisions at $\sqrt{s_{NN}} = 200$
GeV. We also present measurements of mid-rapidity invariant yield at
$\sqrt{s_{NN}} = 62.4$ and $v_{2}$ at $\sqrt{s_{NN}} = 39$ and $62.4$ GeV.
\end{abstract} 

\end{frontmatter} 

\section{Introduction}
Exploiting the merits of heavy quarks is one of the most important and
promising tools to probe the strongly interacting partonic medium created in
heavy-ion collisions. Heavy quarks are mostly created through gluon fusion
\cite{cacciari}, almost exclusively \cite{levai} early in the heavy-ion
collision, therefore, they experience all stages of the medium
evolution. Also, their masses are external to QCD \cite{muller} and thus are
not modified by the presence of the medium. Hence, the kinematics of emerging
heavy quarks carry a memory of their interactions with the medium. By comparing
the heavy quark production in heavy-ion collisions to the baseline production
in $p+p$ and $d$+Au collisions, we seek to further understand the flavor dependence
of energy loss in the medium.  Azimuthal anistropy of heavy quarks
provides further information on the strength of their interaction with the
medium, and better discrimination between theoretical models.

In this article we report on the preliminary results of measuring the
production of electrons from heavy flavor semi-leptonic decays, so-called
non-photonic electrons (NPE).  We show a new high precision measurement of NPE
production at mid-rapidity in \auau\ collisions at \saaTwoHundred, then using
our previously published $p+p$ measurement \cite{starpp200} we show an improved
nuclear modification factor $R_{AA}$ and compare it to theoretical models. Then
we show measurements of NPE azimuthal anistropy, $v_2\{2\}$, $v_2\{4\}$ and
$v_2\{EP\}$. Finally, we show measurements of NPE production in \auau\
collisions at \saaSixtyTwo, and $v_2\{2\}$ measurements at $\sqrt{s_{NN}} = 39$
and $62.4$ GeV.


\section{Datasets and Analyses}

In RHIC run 2010, STAR has sampled nearly 2.6 $nb^{-1}$ luminosity of \auau\
collisions at \saaTwoHundred. Minimum Bias (MB) trigger data is used for
low \pt\  electrons, while high \pt\ trigger (HT) data provides higher statistical
precision for $p_T >$ 2 GeV/$c$. We have also utilized data from an independent $0-5\%$
centrality trigger.  We use about 1 $nb^{-1}$ for the \saaTwoHundred\ results we show here. 
During the same RHIC run STAR has also collected \auau\
collision data at $\sqrt{s_{NN}} = 39$ and $62.4$ GeV, which allows us to search for the onset of heavy quark suppression and flow.

For analyses in all collision energies, the most important detector is the STAR
Time Projection Chamber (TPC) with large acceptance, which provides tracking
and $dE/dx$ for electrons identification.  Hadron rejection is done utilizing
Time of Flight (TOF) \cite{tof} information at low \pt\ and the Barrel
Electromagnetic Calorimeter (BEMC) \cite{bemc} at high \pt.  The BEMC is also used
for triggering on high \pt\ electrons. 

To extract NPE, we statistically subtract the contribution of photonic
electrons (PE) from the identified inclusive electrons.  By reconstructing the
invariant mass of electron pairs we estimate the PE contribution from gamma
conversion, and $\pi^{0}$, $\eta$ Dalitz decays. Extracted yield of PE is then
corrected by a \pt\ dependent reconstruction efficiency determined from
simulation to be at the level of 30-60\%. Other non-photonic background sources are electrons from vector mesons di-electron decays ($p_T <$ 3 GeV/$c$), 
and electrons from heavy quarkonia and Drell-Yan decays ($p_T >$ 2.0 GeV/$c$). Contribution from $J/\psi$ can be non-negligible in this region, 
estimates based on measured $J/\psi$ experimental measurements show it can be as large as 20\% of the measured non-photonic electrons. 
This contribution has been subtracted from \saaTwoHundred\ invariant yields we report in this article.
   
\section{Results}

\begin{figure}[h] \begin{minipage}{.45\textwidth} \vspace*{\fill} \centering
	\includegraphics[width=\textwidth]{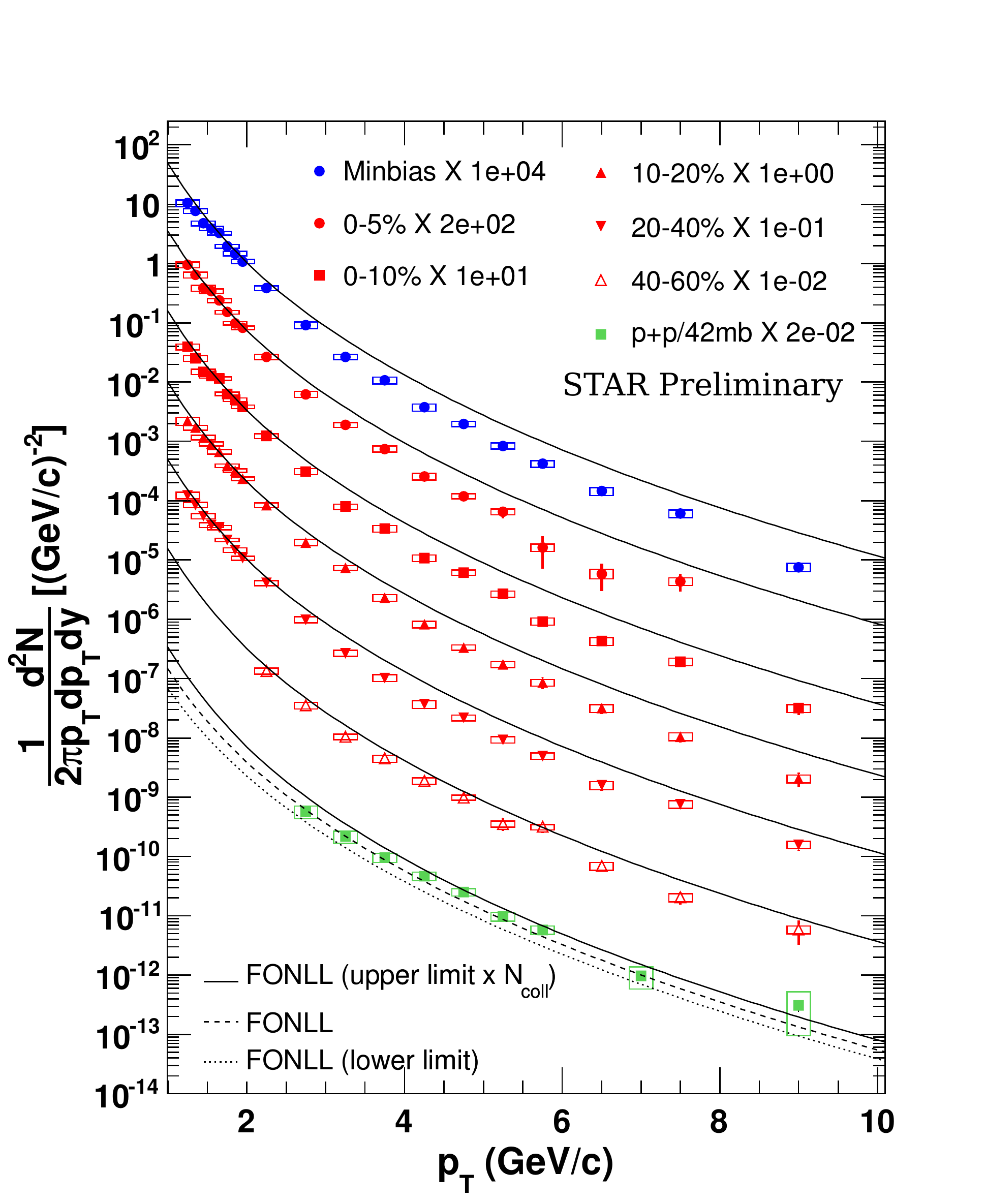} \end{minipage}%
	\begin{minipage}{.4\textwidth} \vspace*{\fill} \centering
		\includegraphics[width=1.07\textwidth]{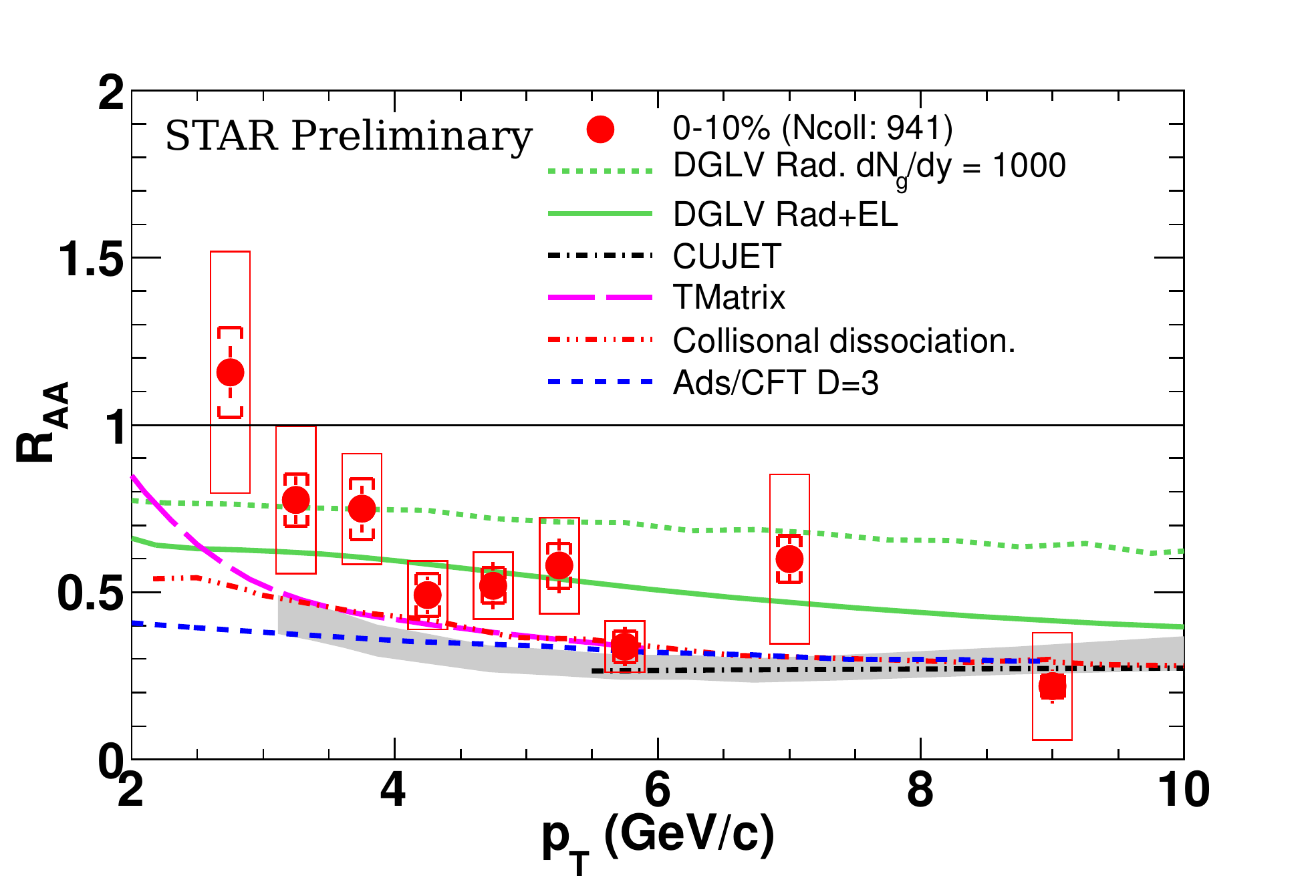}
		\includegraphics[width=\textwidth]{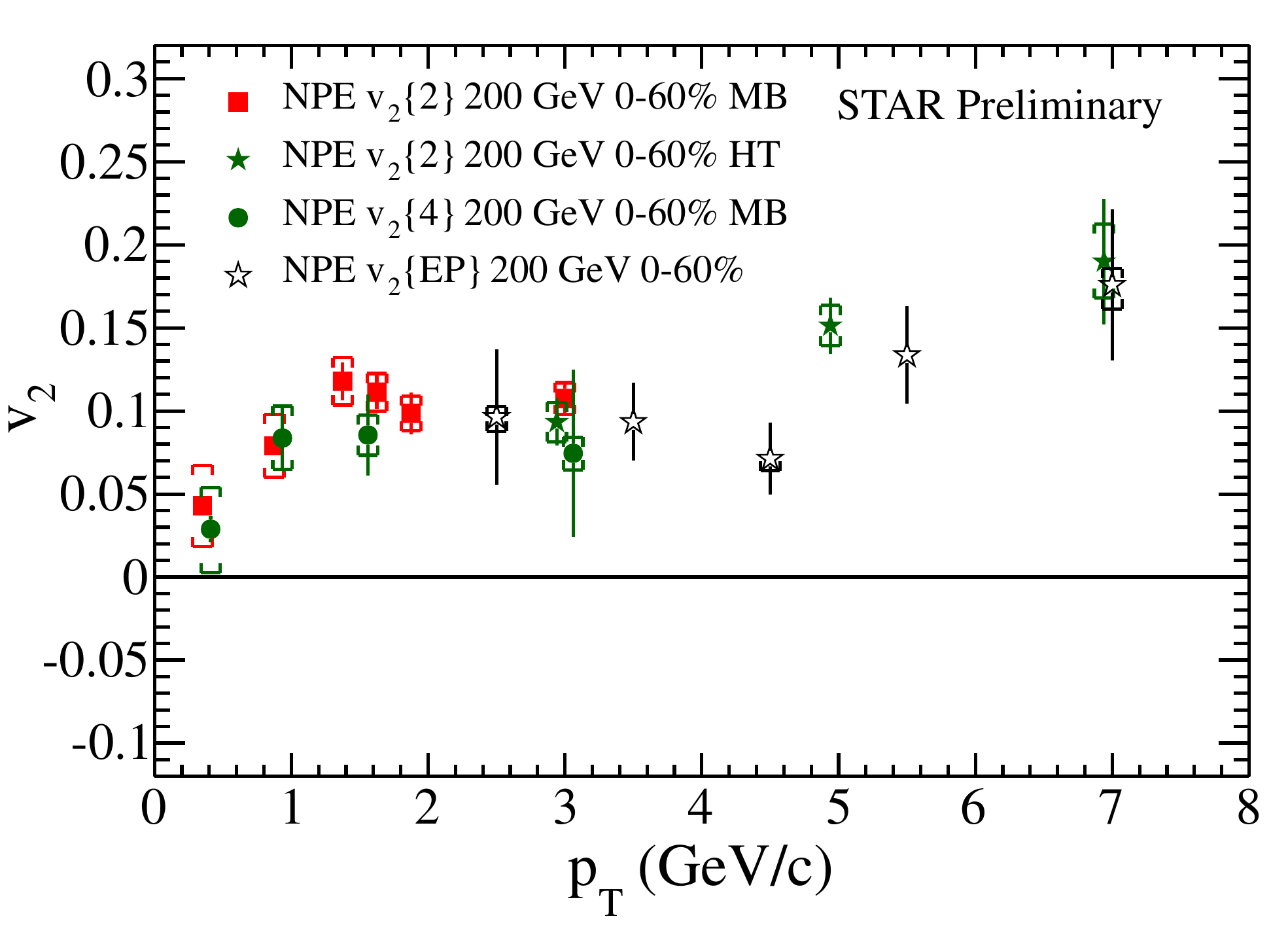}
	\end{minipage} \caption{(Color online) (Left) Invariant yields vs. \pt
		\ of non-photonic electron at \saaTwoHundred, and scaled STAR
		published $p+p$ \cite{starpp200}.  Error bars and boxes are
		statistical and systematic errors, respectively. FONLL
		predictions are scaled by $N_{coll}$ shown as curves.
		(Upper-right) Non-photonic electrons nuclear modification
		factor, $R_{AA}$, at \saaTwoHundred\ compared to models
		\cite{dglv}-\cite{  adscft}. Grey band is the light hadrons
		$R_{AA}$.  Error bars and brackets are Au+Au statistical and
		systematic errors, respectively.  Error boxes are the
		uncertainties from our baseline $p+p$ measurement.
		(Lower-right) Non-photonic electrons azimuthal anistropy
		$v_2\{2\}$, $v_2\{4\}$ and $v\{EP\}$ at \saaTwoHundred. Error
	bars and brackets are statistical and systematic errors, respectively.}
	\label{fig:npe200gev} \end{figure}

%

Figure \ref{fig:npe200gev} left shows the new measurement of NPE mid-rapidity
differential invariant yield for \pt \ = $1.5$ - $10$ \gevc. The novelty of
this measurement lies in the achieved high statistical precision. The large
amount of statistics allows differentiating the measurements in five centrality
bins, in addition to a $0-5\%$ centrality bin from a central trigger. With such
a precision and guided by the scaled FONLL upper bound, one can qualitatively
notice the suppression of the yield in \auau\ collisions compared to $p+p$
collisions despite the large log-scale spanned in the figure.

Figure \ref{fig:npe200gev} upper-right panel shows NPE 
$R_{AA}$ using $0-10\%$ centrality spectra and STAR published $p+p$ \cite{starpp200}
results compared to a collection of models of different energy loss mechanisms.
As we see in the Figure, despite the success of describing the suppression of
light hadrons \cite{dglv}, gluon radiation alone fails to explain the observed
large NPE suppression at high \pt. The large uncertainty from our baseline
$p$+$p$ measurement dominates the current overall uncertainty.  Analysis of the
large amount of collected high quality data from RHIC runs 2009 and 2012 are
needed to improve the baseline precision. However, to exploit the high precision of this measurement, it will be interesting to see theoretical predictions for the quenched invariant yields rather than the $R_{AA}$ only.  With the current precision the
predictions from all other energy energy loss models describe the data.

%
%

\begin{figure}[h] \centering \parbox{2.62in} { \begin{minipage}{.39\textwidth}
	\centering \includegraphics[width=\linewidth]{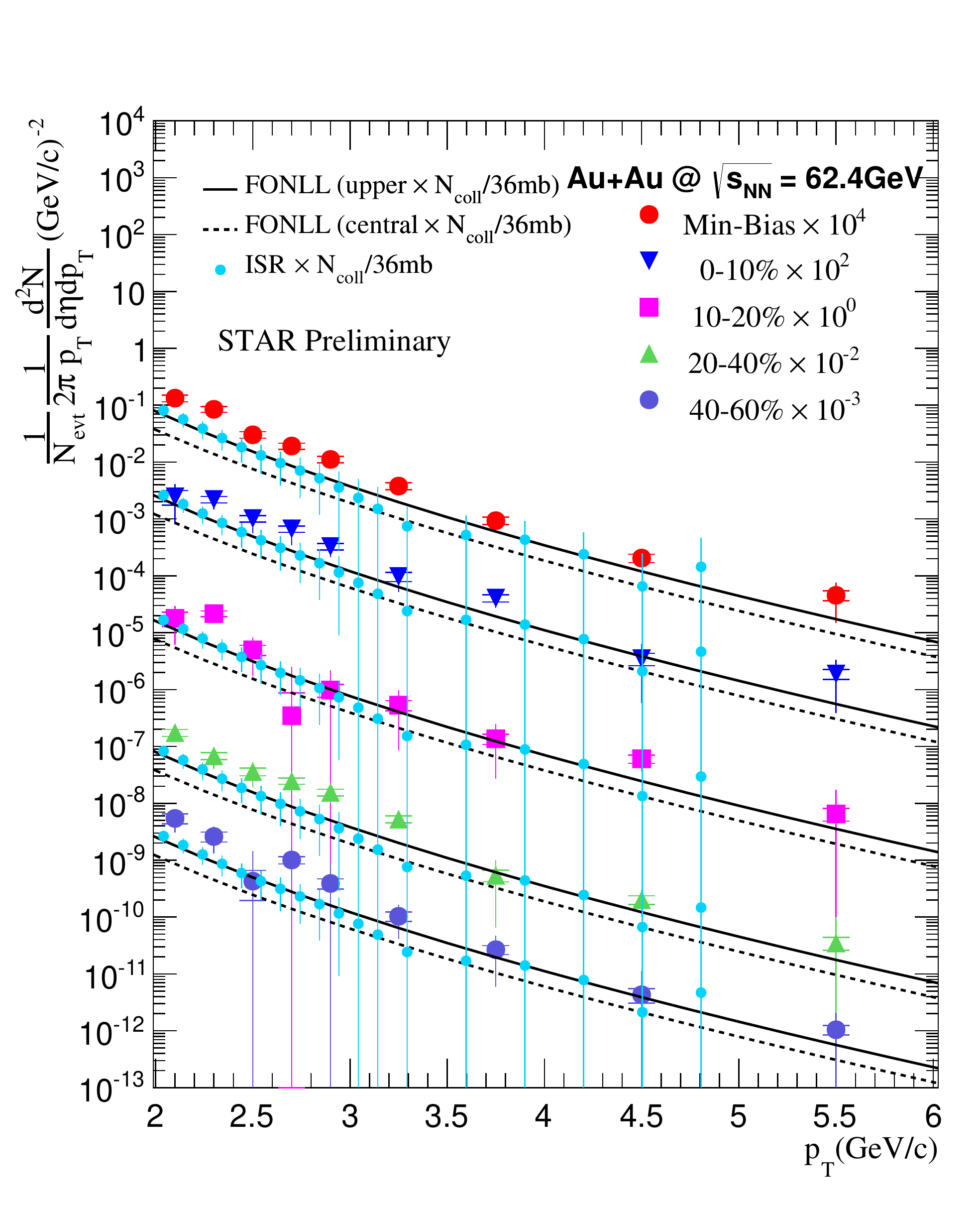}
	\caption{(Color online) Invariant yields vs. \pt \ of non-photonic
		electrons in \auau\ collisions at \saaSixtyTwo. Error bars and
		brackets are statistical and systematic errors, respectively.
		ISR $p+p$ collisions at $\sqrt{s_{NN}}$ = 62.2 GeV scaled by
		$N_{coll}$ is also plotted \cite{isr}.  FONLL predictions are
		scaled by $N_{coll}$ shown as curves.} \label{fig:spec62}
	\end{minipage} } \parbox{2.62in} { \begin{minipage}{.45\textwidth}
		\centering
		\includegraphics[width=\linewidth]{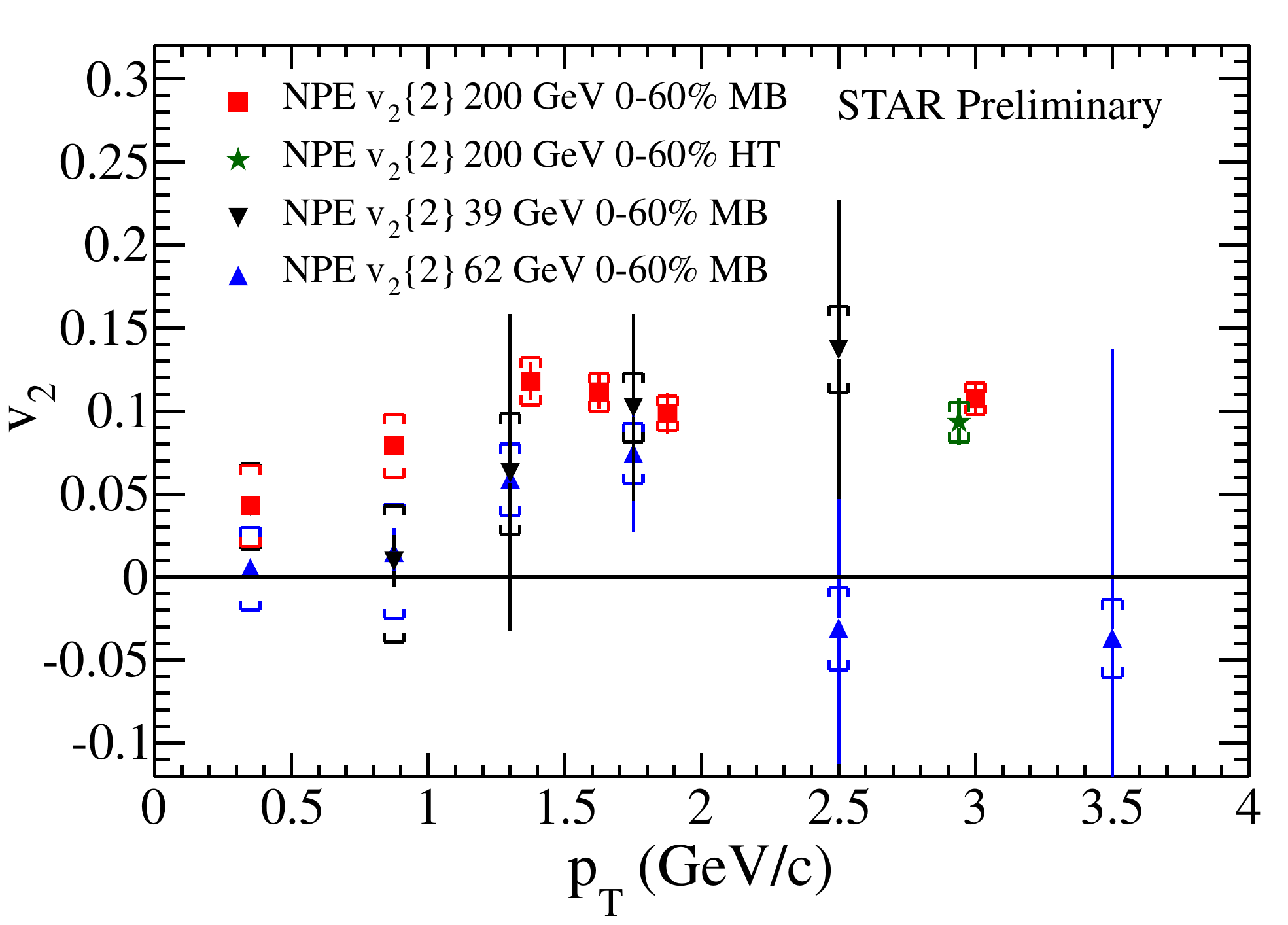}
		\caption{(Color online) Non-photonic electrons at $\sqrt
			{s_{NN}} = 39$, $62.4$ GeV azimuthal anistropy
			$v_2\{2\}$.  Error bars and boxes are statistical and
		systematic errors, respectively.} \label{fig:v2_low_energy}
	\end{minipage} }

\end{figure}

Figure \ref{fig:npe200gev} lower-right shows NPE $v_2$ measurements from 2- and
4- particle correlations and event plane method, represented as $v_2\{2\}$,
$v_2\{4\}$ and $v_2\{EP\}$ in \auau\ collisions at \saaTwoHundred. The $v_2\{2\}$ and
$v_2\{EP\}$ are consistent with each other for $p_T >$ 3 GeV/$c$. While both
show a pronounced systematic increase in $v_2$ towards high \pt, at this point we
cannot distinguish whether this rise is due to jet-like correlations unrelated
to the reaction plane or due to the path length dependence of partonic energy
loss. For \pt\ $< 3$ \gevc\ we show both $v_2\{2\}$ and $v_2\{4\}$. In
$v_2\{4\}$ the non-flow contribution is negligible and the flow fluctuations
contribution is negative, hence providing a lower bound on the $v_2$ of NPE.
Both $v_2$ measurements are finite, which indicates a strong charm-medium
interaction at \saaTwoHundred.

To provide more experimental discrimination power for theoretical models STAR
is extending its NPE program to lower energies. The quest is to see if the
energy loss of heavy quarks is lessened or turned off at lower energies. Figure
\ref{fig:spec62} shows NPE invariant yield in \auau\ collisions
at \saaSixtyTwo\ together with a scaled FONLL prediction. While a previous
$p$+$p$ measurement at the ISR \cite{isr} seems to agree with FONLL upper-band, our
measurement is systematically higher than both. Measurement of $v_2\{2\}$ at
lower energies shown in Figure \ref{fig:v2_low_energy} seem to be consistent
within errors with that at \saaTwoHundred\ for $p_{T} > 1.0$ \gevc.  The results
for data points at $p_{T}<1.0$ \gevc\ seem to hint at a milder charm-medium
interaction compared to those at \saaTwoHundred.

\section{Summary}

In this article we reported on STAR new preliminary results of non-photonic
electron measurements. The new NPE measurements in \saaTwoHundred\ collisions
are precise in a broad \pt\ region. NPE nuclear modification factor measurement
show a large suppression of NPE production in central \auau\ collisions. We
observe large NPE $v_2$ at low \pt\ which indicate a strong charm-medium
interaction. The $v_2$ increases towards higher \pt\ ($>3$ \gevc) is
possibly due to jet-correlations unrelated to the reaction plane and/or due to 
path-length dependence of heavy quark energy loss. 

At lower energies we reported on measurement of NPE invariant
yield in \auau\ collisions at \saaSixtyTwo\ which is systematically higher than
a FONLL prediction. We have also presented our results of azimuthal anistropy
at $\sqrt{s_{NN}} = 39$ and $62.4$ GeV by measuring $v_2\{2\}$ which for
$p_{T}<1.0$ \gevc\ seem to hint at a milder charm-medium interaction than
at \saaTwoHundred.

\end{document}